# Low-Energy Electron-Beam Diagnostics Based on the Optical Transition Radiation

A.N. Aleinik[a], O.V. Chefonov[b], B.N. Kalinin[a], G.A. Naumenko[a*], A.P. Potylitsyn[a], G.A. Saruev[a],
A.F. Sharafutdinov[a], W. Wagner[c]

[a] Nuclear Physic Institute, Tomsk

[b] Russian Federal Nuclear Center. Institute of Technical Physics. Snezhinsk.

[c] Forschungszentrum Rossendorf e.V., Germany

___________________________________________________________________________

**Abstract**

A possibility of a few MeV electron beam diagnostics (Lorentz factor ~ 10) using polarization characteristics of the optical transition radiation (OTR) have been considered. Determinaton of the electron beam divergence have been performed with technique based on the OTR measurements. The technique is based on measuring the ratio $R=I_{min}/I_{max}$, where $I_{min}(I_{max})$ is the photon yield in the minimum (maximum) of OTR angular distribution. It is shown that the value of $R$ can be obtained also from the measurement of the so-called $\psi$-scan of the OTR yield registered by a fixed detector with the aperture $\leq \gamma^{-1}$ by varying the target tilt angle with respect to the electron-beam.

The experiment curried out at 6.1 MeV Tomsk microtron has demonstrated the feasibility of the technique proposed.

[*] Corresponding author E-mail naumenko@npi.tpu.ru

**1. Introduction**

Beam diagnostics based on optical transition radiation (OTR) (see, e.g., [1]) is widely used in electron accelerators within the energy range $10^1$-$10^4$ MeV [2-4]. OTR is commonly employed to measure the transverse beam size and divergence, while coherent transition radiation is used in the millimeter wavelength region to measure a length of short electron bunches [5].

Electron beam divergence can be defined from the shape of the OTR angular distribution (from a single interface or, using the interference methods, from both ones). CCD camera can be used for this purpose. But for low-energy beams (E ~ 10 MeV), the OTR cone becomes quite broad, which decreases the light flux per CCD pixel. With a sensitivity of $10^3$ photons/pixel and the pixel area 100 $\mu m^2$, the optical system converting the OTR angular distribution into a 25 $mm^2$ CCD image ensures an exposure of 0.5 photons per pixel for an electron bunch containing $10^8$ electrons.

This traditional technique is insufficient in this case. However, the usage of a photo multiplier tubes (PMT) more sensitive than CCD camera would provide the required information. Since a PMT measures the total light flux only the information regarding OTR angular distribution may be obtained in the following manner.

The fixed PMT with a narrow angular aperture $\Delta\theta$ detects the OTR yield as dependence on the target inclination angle $\psi$ (so-called $\psi$-scan, see fig.1).

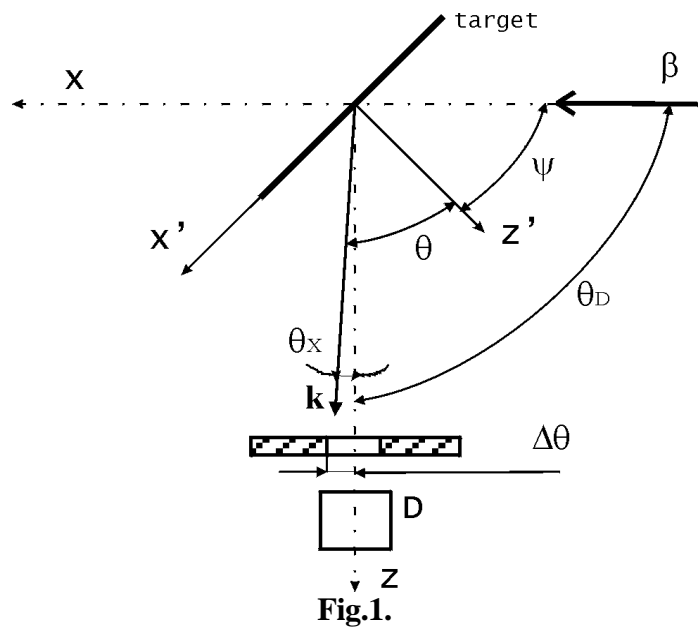

**Fig.1.**

*Coordinate systems describing the OTR process.*
*D is detector, $\boldsymbol{b}$ shows the electron beam direction, $\boldsymbol{k}$ shows the observation direction.*

In the next paragraphs we show that for a small aperture ($\Delta\theta \leq \gamma^{-1}$, $\gamma$ is a Lorenz-factor) the shape of the measured orientation dependence has the same typical features as the OTR angular distribution. It means the beam diagnostics may be based on the measurements of ψ-scans (with polarizing filters upstream PMT too). In this paper we demonstrate the feasibility of proposed scheme.

**2. Theoretical model**

The theory of OTR generated by charged particle crossing an inclined infinite screen has been developed in [6] (for a perfectly conducting screen) and with taking into account the optical properties of screen material in [7]. In the optical range the OTR intensity from such material as polished aluminum differs slightly from ideal screen [8]. So we shall consider the latter case for simplicity.

OTR intensity components polarized in the reflection plane ($I_\parallel$) and in the normal one ($I_\perp$) can be written As [6]:

$$I_\parallel(\vec{n},\omega) = \frac{d^2 W_\parallel(\vec{n},\omega)}{d\omega d\Omega} = \frac{\alpha \beta_{z'}^2 \left|\beta_{x'}\cos\theta_{x'} - \sin^2\theta_{z'}\right|^2}{\pi^2 \left|(1-\beta_{x'}\cos\theta_{x'})^2 - \beta_{z'}^2\cos^2\theta_{z'}\right|^2 \sin^2\theta_{z'}},$$

(1)

$$I_\perp(\vec{n},\omega) = \frac{d^2 W_\perp(\vec{n},\omega)}{d\omega d\Omega} = \frac{\alpha \beta_{x'}^2 \beta_{z'}^2 \cos^2\theta_{y'} \cos^2\theta_{z'}}{\pi^2 \left|(1-\beta_{x'}\cos\theta_{x'})^2 - \beta_{z'}^2\cos^2\theta_{z'}\right|^2 \sin^2\theta_{z'}}.$$

here ω is the OTR photon energy, α is the finite structure constant, $\beta_{x'} = \frac{v}{c}\sin\psi$ and $\beta_{z'} = \frac{v}{c}\cos\psi$ are the charge velocity components in the coordinate system chosen in [6] (see fig.1). In this system the z' axis is perpendicular to the target, and reflection plane is defined as x'z' plane. The OTR-photon outgoing angles are determined by the direction cosines of the wave vector **k** in the same system

$$\cos\theta_{x'} = \sin\theta\cos\varphi,$$
$$\cos\theta_{y'} = \sin\theta\sin\varphi,\quad(2)$$
$$\cos\theta_{z'} = \cos\theta,$$

Where the angles $\theta, \varphi$ are defined in the primed system too.

In the standard geometry for an OTR beam diagnostics photon detectors is placed at right angle relative to an electron beam to detect a backward OTR (see fig. 1). For the reflection plane we have in this case

$$\theta = \frac{\pi}{2} - \psi,$$
$$\varphi = 0$$

(3)

and substituting into Eq.(1) instead cosines Eq.(2) their values from Eg.(3) one may obtain the ψ-scans (orientation dependence of OTR yield):

$$I_{\parallel}(\vec{n}, w) = \frac{b^2 \cos^2 y (\cos y - b \sin y)^2}{(1 - 2b \sin y \cos y)^2},$$

(4)

$$I_{\perp}(\vec{n}, w) = 0,$$

The OTR angular distribution for γ=10 in the reflection plane and for Ψ=π/4 is presented in fig.2 and ψ-scan for θ_D=π/2 at fig.3.

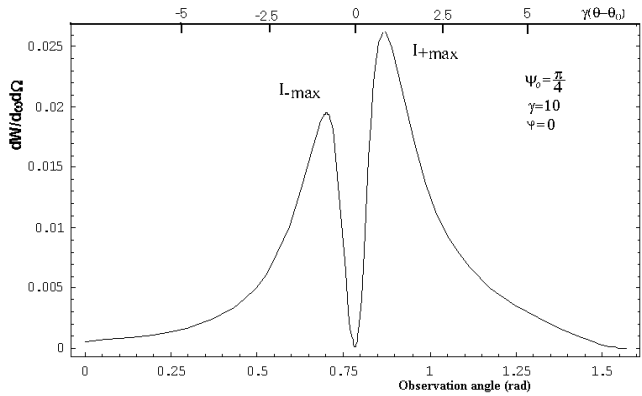

**Fig. 2.**

*Angular distribution of the OTR intensity in the reflection plane; ϕ is the target tilt angle. g - is Lorenz-factor, j is observation angle in plane perpendicular to the speclar reflection plane.*

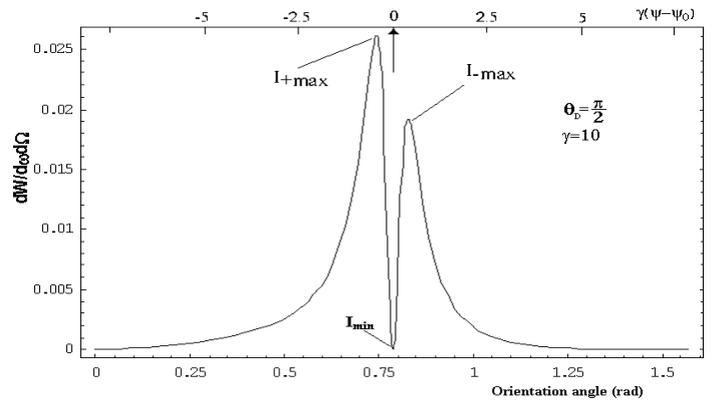

**Fig. 3.**

*Orientation dependence (y-scan) of the OTR intensity in the specular reflection plane; è_D is the observation angle.*

One may see the similarity of both distributions. The difference between right and left parts of distributions ("large shoulder" and "small one") is defined by Lorentz-factor only. Large and small shoulders exchange for ψ-scan due to "reflection" of virtual photons by screen.

The asymmetry ratio is:

$$A = \frac{I_{+\max} - I_{-\max}}{I_{+\max} + I_{-\max}},$$

(5)

where $I_{+\max}$ ($I_{-\max}$) – the maximal value of the OTR intensity for the large (small) shoulder. This value is the same for both figures; A=0.198.

It is evidently (see fig 2) the main part of OTR intensity is concentrated in a cone with apex angle $q \approx g^{-1}$ relative to the specular reflection direction. For a relativistic case (γ>>1) it is more conveniently to use other coordinate system where z-axis coincides with this direction (see fig.1), where the new angular variables may be written as [9]:

$$q_x = q - y,$$
$$q_y = \sin j \sin y, \qquad (6)$$
$$q_x, q_y \approx g^{-1}$$

To calculate the distribution of both polarizations OTR components in the new angular variables (see Fig. 4) the exact expressions (1) can be used:

$$I_\parallel(q_x, q_y) = \frac{ab^2}{p^2} \frac{q_x^2 + q_x(g^{-2} - q_x^2 + q_y^2)tgy}{(g^{-2} + q_x^2 + q_y^2)^2 (1 - q_x tgy)^2},$$

$$I_\perp(q_x, q_y) = \frac{ab^2}{p^2} \frac{q_y^2(1 - q_x tgy)}{(g^{-2} + q_x^2 + q_y^2)^2 (1 - q_x tgy)^2}, \qquad (7)$$

$$I(q_x, q_y) = I_\parallel(q_x, q_y) + I_\perp(q_x, q_y) = \frac{ab^2}{p^2} \frac{q^2 + q_x(g^{-2} - q^2)tgy}{(g^{-2} + q^2)^2 (1 - q_x tgy)^2},$$

$$q^2 = q_x^2 + q_y^2.$$

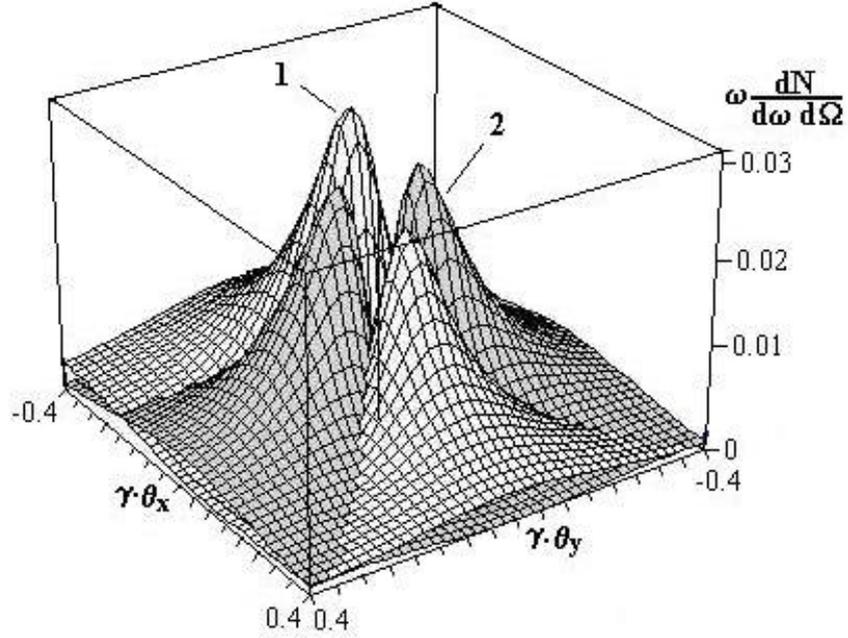

**Fig. 4.**

*OTR angular distribution for both polarization using Eq.(2) for $g=12$ and $y=45^o$.*
*(1 is the angular distribution of the OTR component polarized in the reflection plane, 2 is the same for the perpendicular component).*

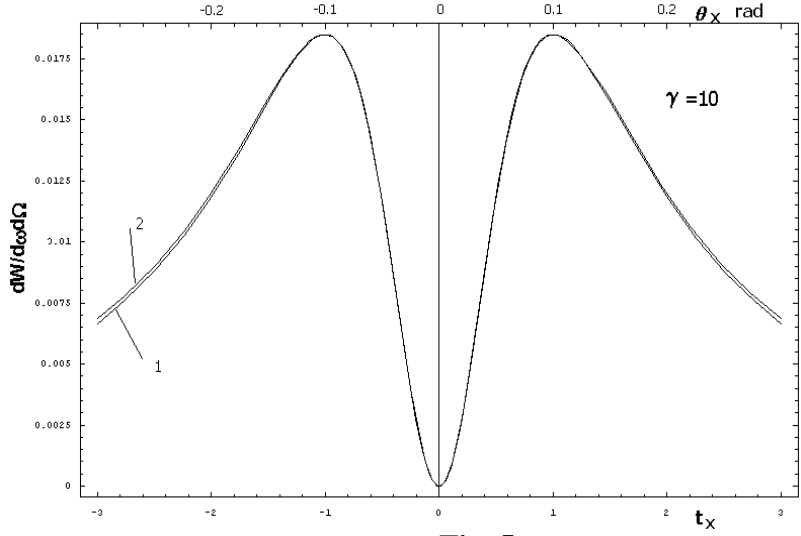

**Fig. 5.**

*OTR angular distribution in the reflection plane. Curve 1 is the calculation using exact formula (1); curve 2 is an approximation being done using Eq. (4). $t_x$ is the observaton angle in units of 1/$\gamma$*

The last expression may be written in more simple form using variables $t_x = \gamma \theta_x, t_y = \gamma \theta_y$ and neglecting by terms $\approx \gamma^{-2}$:

$$\frac{d^2 W}{dt_x dt_y} = \frac{\alpha \gamma^2}{\pi^2}\left[\frac{t^2}{(1+t^2)^2} + \frac{t_x tg\psi}{\gamma(1+t^2)}\right] = \frac{\alpha \gamma^2}{\pi^2}\Lambda(t_x, t_y, \gamma, \psi). \qquad (8)$$

Even for $\gamma$=5 the agreement between approximate equation (8) and exact formula (1) is rather good (see fig. 5) so we use the last expression for calculations with diverging electron beam.

To illustrate our approach let us consider the 2D model where the divergence of the initial beam is described by uniform distribution:

$$F_e(\Delta_x, \Delta_y) = \begin{cases} \dfrac{1}{\pi\sigma^2}, & \Delta_x^2 + \Delta_y^2 \leq \sigma^2 \\ 0, & \Delta_x^2 + \Delta_y^2 > \sigma^2 \end{cases}, \qquad (9)$$

where $\Delta_x$ and $\Delta_y$ are the electron incoming angles in units of 1/$\gamma$, $\sigma$ is a beam divergence parameter in the same system.

To obtain the initial beam divergence effect on resulting OTR angular distribution one should calculate a 2D convolution of distribution (8) and (9)

$$\Lambda_\sigma(t_x, t_y) = \int d\Delta_x \int d\Delta_y F_e(\Delta_x, \Delta_y) \cdot \Lambda(t_x + \Delta_x, t_y + \Delta_y). \qquad (10)$$

The result of this convolution for the beam divergence $\sigma$=0.5 is shown for both components on fig.6 and fig.7.

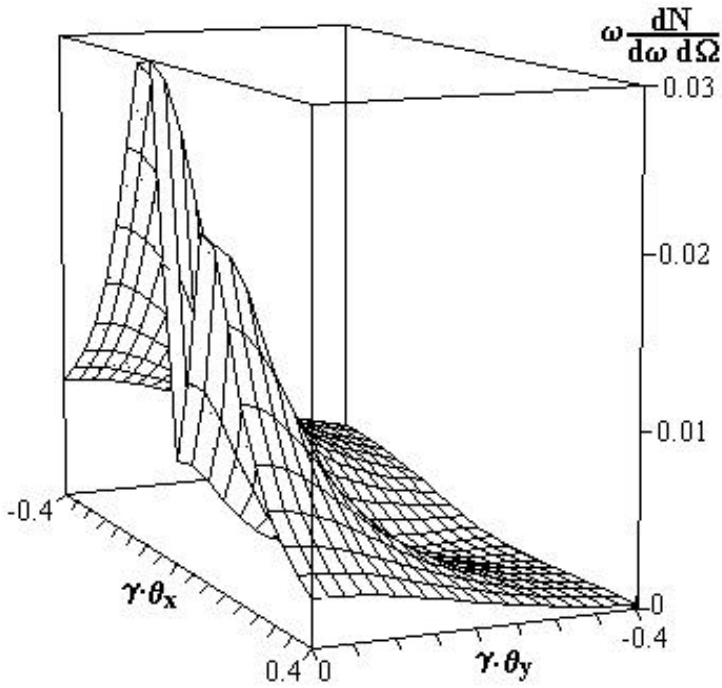

**Fig. 6.**

*Angular distribution of the OTR parallel component calculated using Eq.(1) for the electron beam divergence with $s=1/2\,g$*

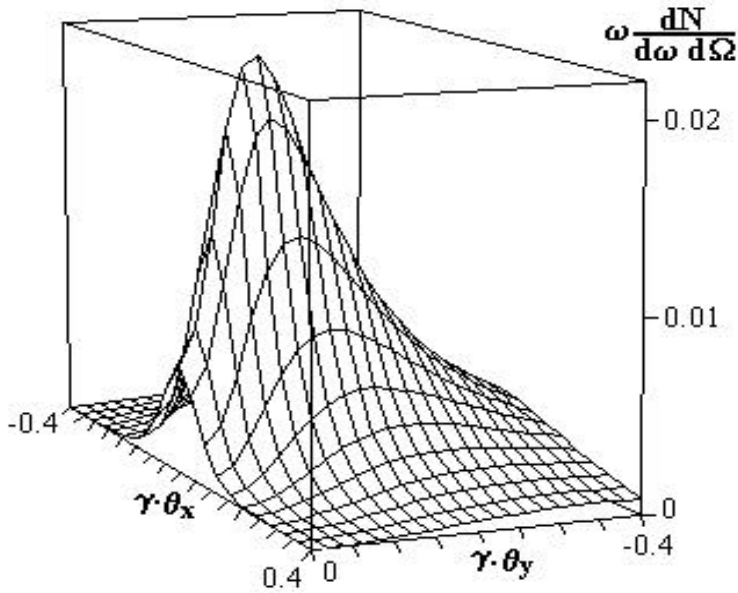

**Fig. 7.**

*Angular distribution of the OTR perpendicular component calculated using Eq.(1) for the electron beam divergence with $s=1/2\,g$*

From Fig.7 it can be seen that there is the significant contribution from the perpendicular polarized component. Other effect from diverging electron beam is the "smoothed" nonzero minimum along specular reflection direction (compare fig.2 and 6).

The "contrast" between minimal and maximal values of OTR intensity may characterize the beam divergence effect. Let's introduce the following criterion:

$$R = \frac{I_{\min}}{\frac{1}{2}(I_{+\max} + I_{-\max})}. \qquad (11)$$

We assume the denominator behavior may be approximated by the first term in Eq.(8):

$$\frac{1}{2}[I_+(t_X, t_Y = 0) + I_-(t_X, t_Y = 0)] = \frac{a g^2}{p^2} \frac{t_X^2}{(1+t_X^2)^2} = \frac{a g^2}{p^2} \Lambda^\circ(t_X, t_Y = 0). \qquad (12)$$

The distribution $\Lambda°(t_X, t_Y = 0)$ doesn't depend on Lorenz-factor and describes the OTR angular distribution for perpendicular screen in the ultrarelativistic case. It may be shown readily the equation (12) is fulfilled exactly for a non-diverging beam (σ=0). So we suppose the accuracy of this approximation will remain good for divergences $\textbf{\textit{s}} \leq 1$.

The 2D integral Eq.(10) is calculated analytically for our model electron divergence (9) using the polar coordinate system and can be reduced to the following:

$$\Lambda_s(t_x, t_y) = \frac{1}{2\textbf{\textit{s}}^2}\left\{\frac{1+t^2-\textbf{\textit{s}}^2}{\sqrt{D}} - 1 + 2\ln\frac{1+\textbf{\textit{s}}^2-t^2+\sqrt{D}}{2}\right\}, \qquad (13)$$

where $D = (1+t^2)^2 + 2\textbf{\textit{s}}^2 \cdot (1-t^2) + \textbf{\textit{s}}^4$, $t^2 = t_x^2 + t_y^2$.

The resulting expression is considerably simplified for the case of importance ó<< 1. Expanding Eq. (13) with respect the power of the parameter σ and leaving the terms ó$^2$, we obtain

$$\Lambda_s(t_x, t_y) = \frac{1}{(1+t^2)^2}\left[t^2 + \textbf{\textit{s}}^2\left(\frac{1}{2} - \frac{3t^2}{(1+t^2)^2}\right)\right]. \qquad (14)$$

The same folding procedure allows to obtain the parallel component of the resulting OTR intensity

$$\Lambda_{s\|}(t_x, t_y) = \frac{1}{2\textbf{\textit{s}}^2}\left\{\frac{t_x^2}{t^2}\left[\frac{1+t^2+\textbf{\textit{s}}^2}{\sqrt{D}} - 1\right] + \frac{t_x^2 - t_y^2}{2t^4}\left[1 + \textbf{\textit{s}}^2 + t^2 - \sqrt{D}\right] + \ln\frac{1+\textbf{\textit{s}}^2-t^2+\sqrt{D}}{2}\right\}. \qquad (15)$$

The expression for perpendicular component is derived from (15) after substitution $t_x \leftrightarrow t_y$. In analogy with Eq.(14) we have more simple expressions for a low beam divergence:

$$\Lambda_{s,\|,\perp}(t_x, t_y) = \frac{1}{(1+t^2)^2}\left[t_{x,y}^2 + \textbf{\textit{s}}^2\left(\frac{1}{4} - \frac{3t_{x,y}^2}{(1+t^2)^2}\right)\right]. \qquad (16)$$

It is obvious that the results obtained are valid for the detector with an infinitely small aperture. Since to measure the coefficient $R$ (from the measured ψ-scan) use is made of a detector with a finite aperture, it is necessary to run addition-folding procedure. To calculate the OTR yield in the circular detector aperture with an angular size acceptance $\Delta\theta = \textbf{\textit{d}}$ (see fig. 1), we have to integrate Eq. (13) with respect to the aperture

$$Y(t_x, t_y) = \int_{\Delta\Omega} d\textbf{\textit{d}}_x d\textbf{\textit{d}}_y \cdot \Lambda_s(t_x + \textbf{\textit{d}}_x, t_y + \textbf{\textit{d}}_x), \qquad (17)$$

where $\delta_x^2 + \delta_y^2 < \delta^2$, $t_x$ and $t_y$ are the angular coordinates of the aperture center.

For case δ<< 1, we are attracted, the approximate formula can be also obtained. Expanding Eq. (17) in powers of the parameter δ and leaving the terms not higher than $\delta^4$, $\sigma^4$, we obtain taken into account Eq. (14):

$$R \approx \frac{Y(0,0)}{Y(1,0)} = \frac{8(3(s^2 + d^2) - 4s^4 - 4d^2 \cdot (3s^2 + d^2))}{3(4 - (s^2 + d^2))}. \qquad (18)$$

Leaving in Eq. (18) the first non-zeroth terms, we obtain

$$R \approx 2(d^2 + s^2). \qquad (19)$$

Let us remind that $\sigma$ is the divergence of electron beam, but $\delta$ is the angular aperture of the detector. Using the same convolution procedure and leaving in $Y(t_x, t_y)$ the second orders terms, we may derive the criterion of contrast for the OTR component polarized in the reflection plane for circular detector aperture:

$$R_\| \approx \frac{2(3(s^2 + d^2) - 4s^4 - 4d^2 \cdot (3s^2 + d^2))}{3(2 - (s^2 + d^2))} \approx s^2 + d^2. \qquad (20)$$

Now let us consider the case of an electron beam with different divergences $\sigma_x$ and $\sigma_y$ ($\sigma_y \ll \sigma_x = \sigma$, the case of the "flat" electron beam). The angular OTR distribution will result in this case by the following convolution:

$$\Lambda_s^-(t_x, t_y) = \int F_e(\Delta_x, 0) \cdot \Lambda(t_x + \Delta_x, t_y) \cdot d\Delta_x. \qquad (21)$$

Using the similar expansion procedure, we can derive the criterion of contrast for the both components of radiation for this case:

$$R^- \approx \frac{8}{5} \cdot \frac{2s^2(5 - 6s^2) + 5d^2(3 - 8s^2) + 20d^4}{12 - 4s^2 - 3d^2} \approx \frac{4}{3}s^2 + 2d^2. \qquad (22)$$

Lastly, using the similar way, we may derive the criterion of contrast for the OTR component polarized in the reflection plane for the case of experimental conditions $\sigma_y \ll \sigma_x = \sigma$:

$$R_\|^- \approx \frac{2}{5} \cdot \frac{4s^2(5 - 6s^2) + 5d^2(3 - 14d^2) - 20d^4}{2(3 - s^2) - 3d^2} \approx \frac{4}{3}s^2 + d^2. \qquad (23)$$

Analyze of these results for the diverging beam described by the uniform distribution (see Eq. (16) shows, that the contrast-range coefficient R for the OTR component polarized in the reflection plane, as a function of $\sigma$, is different from the non-polarized case by a factor of ½ and it is the same in comparison to the detector aperture contribution. So the usage of a polarizing filter for this case is not effective. However, for the flat electron beam (see Eq. (22,23) the effect of beam divergence in comparison to the detector aperture contribution is larger by a factor of 2. Hence the usage of polarized OTR component for diagnostics is preferable in this case.

Having these results we can find the beam divergence from experimental data. Parameter $ó$ may be determined from the measured value of R using relations (19, 23):

From Eq.(12) for the azimuthally uniform distribution of electron beam divergence we have

$$\mathbf{s} = \sqrt{\frac{R}{2} - \mathbf{d}^2} \;, \qquad (24)$$

and for the flat electron beam

$$\mathbf{s} = \frac{\sqrt{3}}{2} \cdot \sqrt{R - \mathbf{d}^2} \qquad (25)$$

## 3. Experiment

The experimental study of OTR dependence characteristics on electron beam parameters were performed at the microtron of NPI TPU. The main designed microtron parameters are as follows:

| | |
|---|---|
| Electron energy | 6.1 MeV |
| Macropulse duration | 2 – 6 µs |
| Pulse repetition rate | 1 – 10 Hz |
| Micropulse duration | 17 – 20 ps |
| Current amplitude per macropulse | 40 mA |
| Micropulse length | 0.6 cm |
| Number of micropulses per macropulse | $10^4$ |
| Current amplitude per micropulse | 0.6 A |
| Average microtron current (at 10 Hz) | 2.4 µA ($1.5 \cdot 10^{13}$ ç/s) |
| Energy variance | 0.5% |
| Beam size at the microtron output | 4 ~2 mm$^2$ |
| Emittance: horizontal | $3 \cdot 10^{-2}$ mm·rad |
| vertical | $1.5 \cdot 10^{-2}$ mm·rad |

The beam parameters allow us to consider the electron beam as a flat one and we can use the formulas (22,23,25).

The experimental layout is shown in Fig.8. Four quadruple lenses ensure the electron beam focusing on the target. The beam size can be monitored with a video camera from the scintillation-counter display and, to obtain a more precise vertical beam size – with a wire scanner.

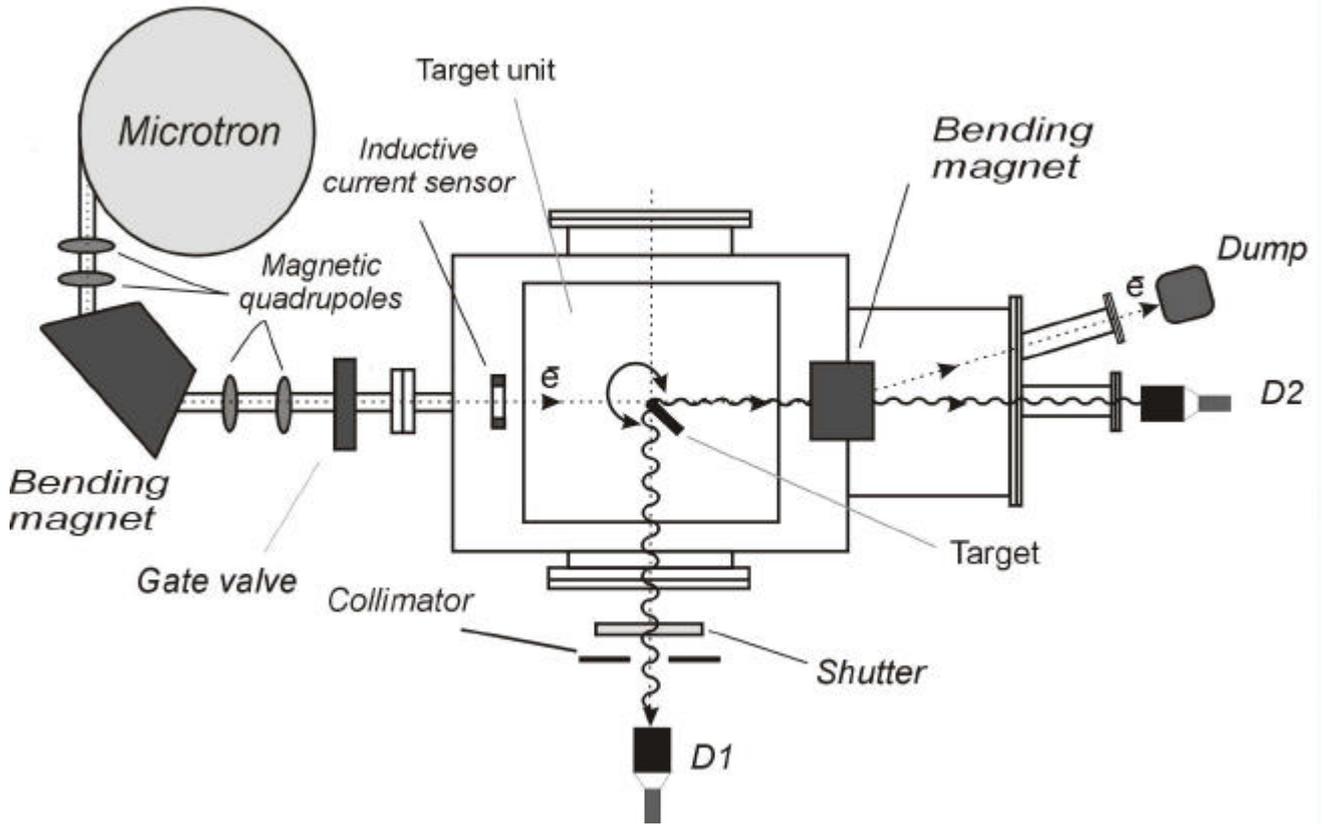

**Fig. 8.**

*Experimental layout. D1 is optical detector, D2 is g-detector.*

OTR target was made from a thin aluminum mirror deposited onto a ceramic substrate. Their thickness was 1ìm and 0.22 mm, respectively. The target was fixed in the goniometer and could be rotated using a step motor. The electron current and the forward bremsstrahlung yield from the target were also measured. In order to take the radiation background into account, the latter was measured for each target position with the shutter closing the entrance window of PMT. All the measurements were normalized with respect to the electron beam current. Magnetic lenses used in the experiment made it possible to vary the beam size focused on the target. Following the beam focusing, the orientation characteristics of transition radiation were measured. For a fixed beam emittance, the angular divergence decreases with increasing the linear beam size.

OTR was measured using a PMT located at 90° to the beam direction at the distance of 1200 mm from the target. The circular collimator placed on the detector was 20 mm in diameter, which corresponds to the aperture $d = 0.1$ in units of $1/\gamma$. Fig.9 and Fig.10 present the measured curves of the TR orientation dependence in terms of $dW/d\Omega$ (sum of both polarization components) for different electron beam sizes $\Delta$ (in other words for different beam divergences). It is evident from the figures that the beam divergence severely affects the orientation dependence shape.

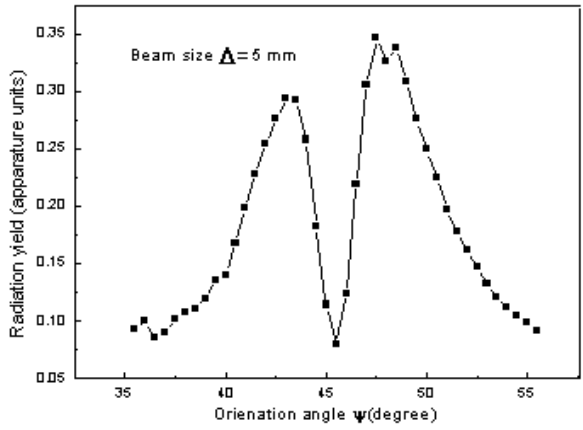 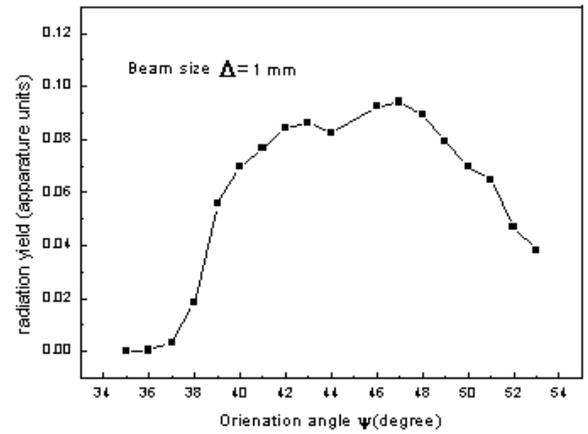

**Fig. 9.**

*Experimental **y**-scan for the beam size **D** = 5 mm. Solid line is the interpolation of experimental data*

**Fig. 10.**

*Experimental **y**-scan for the beam size **D** = 1 mm. Solid line is the interpolation of experimental data*

## 4. OTR polarization and its application in electron beam diagnostics

The optical analyzer allowed us to measure the yield of the OTR components polarized parallel $(I_\parallel)$ and perpendicular $(I_\perp)$ to the reflection plane. Since the spectrum of OTR components is the same in the visible range, we believe that the instrument asymmetry associated with the rotation of the analyzer by 90 is negligible. Figure 11 depicts the measured ψ-scan from different components of the OTR intensity for the vertical beam size Δ=2.6 mm.

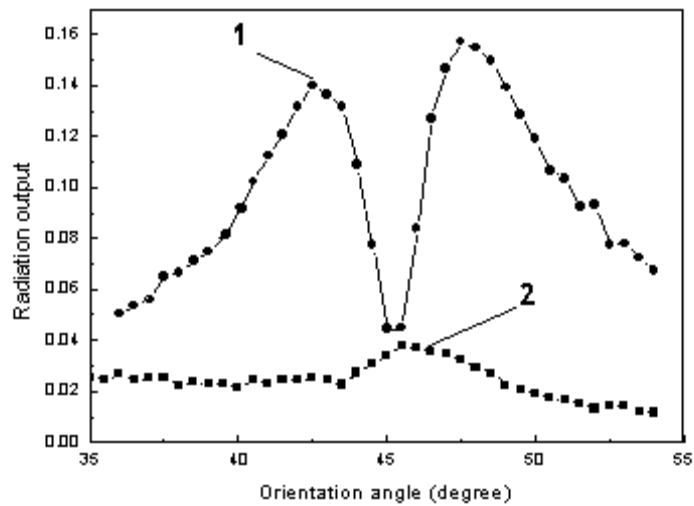

**Fig. 11.**

*Experimental **y**-scan of both polarization components for the beam size **D**=2.6 mm. Curve 1 is the orientation dependence of the OTR component polarized in the reflection plane, curve 2 is the same for the perpendicular component. Solid lines are the interpolations of experimental data.*

The figure shows a principle agreement with theory for both polarization components (see Figs. 6, 7). Similar dependences were measured for several values of Δ.

Shown in Fig.12 are the results of measuring the contrast-range coefficient $R(\Delta)$ as a function of beam size $\Delta$ in x-direction. Also using Eq. (25) the electron beam divergence for each size of electron beam was from experimental data recovered (Fig. 13).

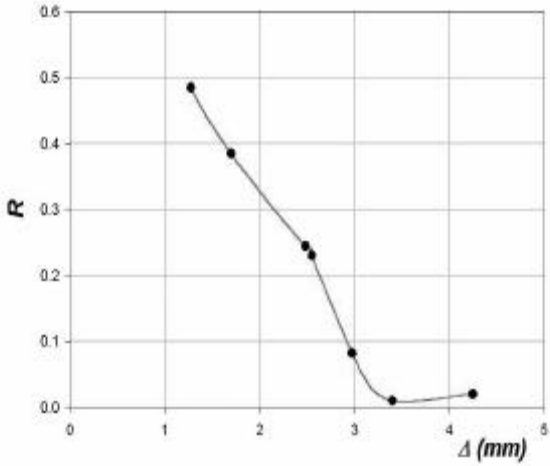

**Fig. 12.**

*Experimental dependence of the contrast-range coefficient on a beam size. Solid line is the interpolation of experimental data.*

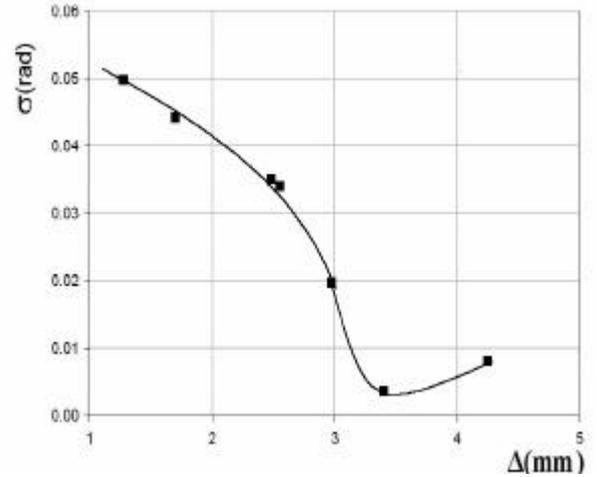

**Fig. 13.**

*Beam divergence as a function of a beam size recovered from the experimental data.*

**Summary**

The possibility of low-energy electron beam diagnostics using the polarization characteristics of optical transition radiation (OTR) has been considered. The OTR yield $\psi$-scans has been offered as a tool to investigate the beam characteristics. Calculations have shown that it is the radiation polarized in the reflection plane (parallel component) that is of the greatest interest for diagnostics. The orientation dependence of this component has certain peaks and dips that disappear with an increase in angular divergence of the initial electron beam. The ratio R of the radiation yield in the dip to half sum of the maxima has a unique dependence on angular divergence and can be used for measurement of one. The experimental study of the OTR component in reflection plane was carried out at the microtron of the Nuclear Physics Institute with electron energy 6.1 MeV, the orientation dependence. It has been shown the *R*-ratio depends on beam size, which can be used to estimate beam divergence. Also, the parallel and perpendicular components of OTR are given. The nonzero contribution from the perpendicular component along the specular direction is attributed to beam divergence and the geometry of the experiment. Similar patterns were obtained in the calculations taking into account the experimental geometry. Unfortunately, only a wire scanner was available in the experiment to verify the beam size, and no means for the angular beam divergence. Nevertheless, the proposed technique allows us to analyze the beam divergence from ratio R that may be measured from $\psi$-scan easily.

In Ref. [10] it was noted that determination of a beam divergence from the shape of OTR angular pattern for 1MeV electrons is ineffective.

We have demonstrated experimentally that ψ-scans can provide the same information regarding such typically measured features as the shape of its angular pattern. The technique offered can be used to an advantage for the beam intensity as low as $10^8$ $\bar{e}/s$ due to high sensitivity of the detector used (PMT).

**Referecies**